\documentclass[sigconf,dvipsnames]{cidr-2026}

\AtBeginDocument{%
  }

\usepackage{multirow}
\usepackage{graphicx}
\usepackage{subcaption}
\usepackage{tikz}
\usepackage{xcolor}
\usepackage{amsthm}
\usepackage{wrapfig}
\usepackage{enumitem}
\usepackage{pifont}
\usepackage{algpseudocode}
\usepackage{mdframed}
\usepackage{hyperref}[colorlinks=true, linkcolor=blue, urlcolor=blue, citecolor=blue]
\usepackage[utf8]{inputenc}
\usepackage[most]{tcolorbox}
\usepackage[labelfont=bf]{caption}
\usepackage{breqn}
\usepackage{float}
\usepackage{amsmath}
\usepackage{changepage}
\usepackage{dsfont}
\usepackage{adjustbox}
\usepackage{fancyvrb} 
\usepackage{minted}
\usepackage[vlined,commentsnumbered,ruled]{algorithm2e}
\usepackage[table]{xcolor}
\usepackage{listings}

\definecolor{codegray}{gray}{0.95}
\lstdefinestyle{mypython}{
    backgroundcolor=\color{white},
    basicstyle=\ttfamily\footnotesize,
    keywordstyle=\color{blue},
    commentstyle=\color{gray},
    stringstyle=\color{orange},
    showstringspaces=false,
    breaklines=true,
numbers=left,
numbersep=5pt,
xleftmargin=2em,
framexleftmargin=2em,
    language=Python
}

\usepackage{array}          %
\usepackage{makecell}       %

\newcolumntype{C}[1]{>{\centering\arraybackslash}m{#1}} %

\tcbuselibrary{listings}      %
\definecolor{promptgray}{RGB}{245,245,245}   %
\definecolor{promptblue}{RGB}{235,245,255}    %
\definecolor{promptyellow}{RGB}{255,255,220}  %
\definecolor{promptorange}{RGB}{255,240,210}  %

\newtheoremstyle{no-indent-theorem}  %
  {}  %
  {}  %
  {\itshape}  %
  {0pt}  %
  {\scshape}  %
  {.}  %
  { }  %
  {}  %

\theoremstyle{no-indent-theorem}

\newtheoremstyle{exampleStyle}%
  {3pt}   %
  {3pt}   %
  {}      %
  {}      %
  {\bfseries} %
  {.}     %
  { }     %
  {}      %
\theoremstyle{exampleStyle}

\newcommand*\circled[1]{\tikz[baseline=(char.base)]{
            \node[shape=circle,draw,inner sep=1pt,fill=black,text=white] (char) {#1};}}

\usepackage{xspace}

\newcommand{\sys}{\textsc{Buckaroo}\xspace}
\usepackage{fontawesome}

\usepackage{tikz}

\newcommand{\parahead}[1]{\medskip
\noindent\textbf{#1}}

\newcommand{\custombullet}[2]{%
  \noindent\scalebox{0.6}{  \ding{108}  }%
  \ifx&#1&%
    \textbf{ }#2%
  \else%
    \textbf{#1: }#2%
  \fi%
}

\newcommand{\customsection}[1]{%
    \vspace{0.5em} %
    \noindent
    \ifx&#1&%
        \textbf{ }%
    \else%
        \textbf{#1. }%
    \fi%
}

\usepackage{array}

\usepackage[T1]{fontenc}
\usepackage{seqsplit}
\usepackage{xcolor} %

\begin{document}

\title{Towards Scalable Visual Data Wrangling via Direct Manipulation}

\author{El Kindi Rezig}
\affiliation{%
  \institution{University of Utah}
  \country{}
  }
\email{elkindi.rezig@utah.edu}

\author{Mir Mahathir Mohammad}
\affiliation{%
 \institution{University of Utah}
 \country{}
 }
\email{mahathir.mohammad@utah.edu}

\author{Nicolas Baret}
\affiliation{%
  \institution{University of Utah}
  \country{}
  }
\email{nicolas.baret@utah.edu}

\author{Ricardo Mayerhofer
 }
\affiliation{%
  \institution{Hopara, Inc}
  \country{}
  }
\email{ricardo@hopara.io}

\author{Andrew McNutt}
\affiliation{%
  \institution{University of Utah}
  \country{}
  }
\email{andrew.mcnutt@utah.edu}

\author{Paul Rosen}
\affiliation{%
  \institution{University of Utah}
  \country{}
  }
\email{paul.rosen@utah.edu}

\begin{abstract}
Data wrangling—the process of cleaning, transforming, and preparing data for analysis—is a well-known bottleneck in data science workflows. A wide range of data wrangling techniques have been proposed to mitigate this challenge. Of particular interest are visual data wrangling tools, in which users prepare data  via graphical interactions (such as with visualizations) rather than requiring them to write scripts. 
We develop a visual data wrangling system, \sys that expands upon this paradigm by enabling the automatic discovery of interesting groups
(e.g., Salary values for Country=``Buthan'') and identification of anomalies (e.g., missing values, outliers, and type mismatches) both within and across these groups. Crucially, this allows users to reason about how repairs applied to one group affect other groups in the dataset.

A central challenge in visual data wrangling is scalability. Rendering entire datasets is often infeasible, yet showing only a small sample risks hiding rare but critical errors across groups.
We address these challenges through 
carefully designed sampling strategies that prioritize errors, as well as novel aggregation techniques that support pan-and-zoom interactions over large datasets.
\sys maintains efficient indexing data structures and differential storage to localize anomaly detection and minimize recomputation. 
We demonstrate the applicability of our approach via an integration with the \textit{Hopara} pan-and-zoom engine (enabling multi-layered navigation over large datasets without sacrificing interactivity). Finally, we explore our system's usability (via an expert review) and its scalability, finding that this design seems well matched with the challenges of this domain.\footnote{This version revises the paper to improve its positioning with respect to prior work in response to feedback received after publication. Earlier versions remain available in the arXiv version history.}  
\end{abstract}

\maketitle

\section{introduction}

The promise of data-driven decision-making relies critically on the quality and readiness of underlying datasets~\cite{gartner}. Yet, before any analysis, modeling, or visualization can occur, practitioners must invest substantial effort into \emph{data wrangling}---the process of transforming raw, messy data into a structured form suitable for downstream tasks. 
Despite its importance, data wrangling remains one of the most labor-intensive and error-prone phases of the data science lifecycle, accounting for up to 80\% of the total project time~\cite{elkindicidr}.

Data wrangling involves a wide range of tasks, including parsing, deduplication, missing value imputation, anomaly detection, and type normalization. 
Further complicating these challenges is that fixing one data anomaly can lead to other anomalies. Several approaches have been proposed to mitigate error detection and correction~\cite{readytodeploy}. In particular, \textit{visual} data wrangling techniques~\cite{kandel2011wrangler, kandel2012profiler} enable users to identify and correct data anomalies through graphical interactions such as visualizations. We present \sys, which extends this paradigm by introducing three key capabilities for visual data wrangling: (1)~\textbf{Group-level anomaly detection}, where \sys automatically generates \textit{interesting} data groups (e.g., salary values for Country = ``Bhutan'') and detects anomalies both within and across groups (e.g., a value may be an outlier in one group but not in another); (2)~a \textbf{pan-and-zoom interface} that allows users to focus on regions of interest within a visualization; and (3)~\textbf{anomaly-centric sampling and aggregation}, which provides multiple sampling and aggregation strategies to support scalable visual data wrangling on large datasets.

Data wrangling for anomaly detection is rarely a one-shot operation—particularly when anomalies are \emph{subgroup-dependent}. An error that appears anomalous in one subset of the data may be expected or even valid in another, necessitating iterative, context-aware inspection and correction informed by domain knowledge. In this work, we propose to (1) identify semantically meaningful data groups (e.g., \textit{Education} for \textit{Country} = “Bhutan”), (2) detect anomalies within and across these groups, and (3) support their repair through scalable visual interfaces.

It is well known that visualization tools excel at surfacing structure in data~\cite{vistrails, ruddle2023tasks} and supporting data wrangling tasks~\cite{kandel2011wrangler, kandel2012profiler}. We build on this line of work by enabling data wrangling for \emph{subgroup} anomaly detection and correction through interactions with interactive charts. Our proposed system, \sys, frames subgroup data wrangling as a visual, interactive, and iterative process.
It allows users to identify and resolve group-level data anomalies by directly manipulating visual representations of data groups. \sys automatically detects anomalous groups---e.g., those with missing values or outliers---maps them to interactive charts, and offers recommended repairs that can be applied, visualized, and reverted in real time. This tight integration of detection, visualization, and repair enables users to understand the impact of each action across the dataset and supports the inherently exploratory nature of data preparation.

\begin{figure*}[htbp]
    \centering
        \includegraphics[width=\linewidth]{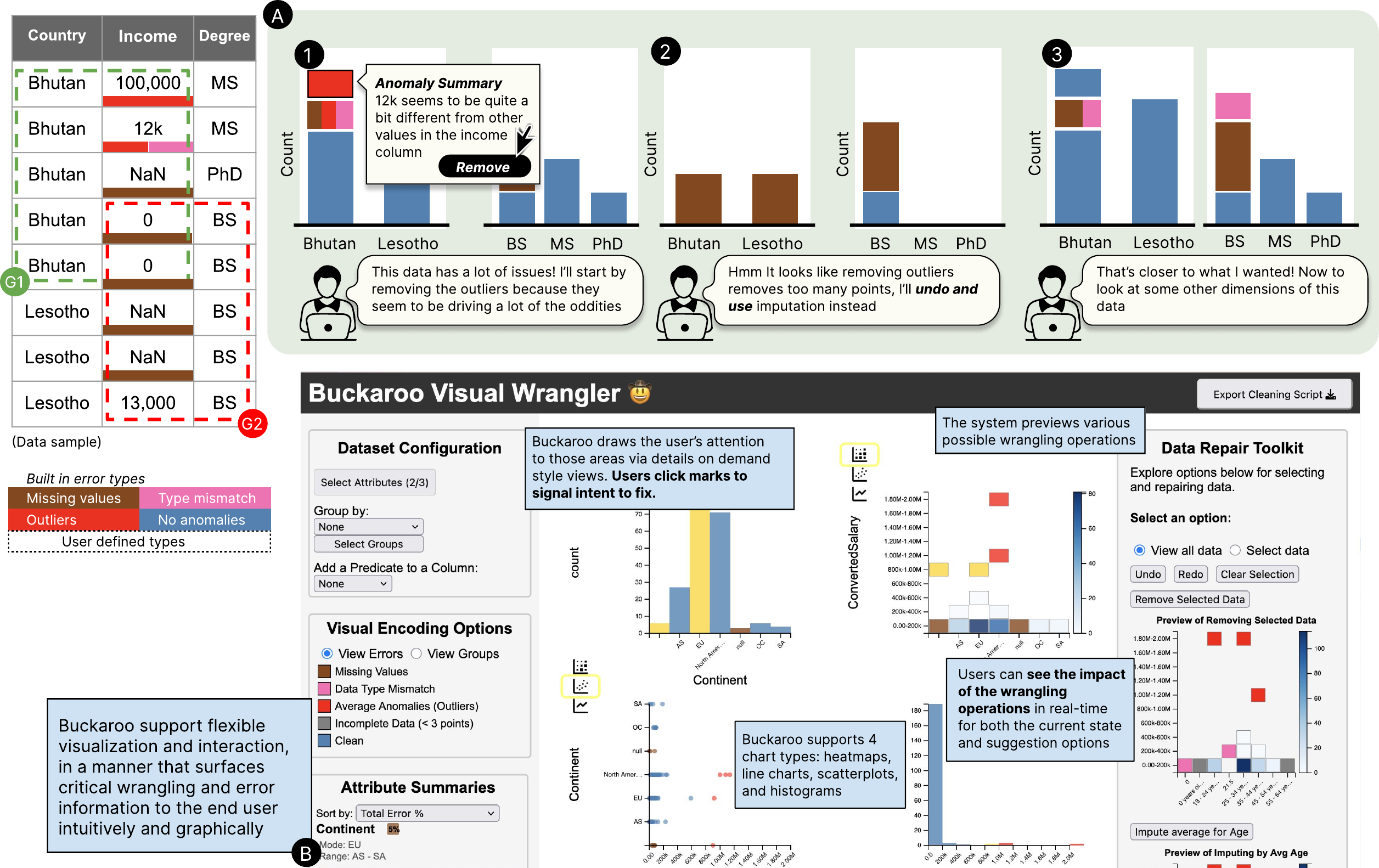}
    \caption{An overview of repairing an error through \sys's user interface. \protect\circled{A} highlights a user working through iterative and backtrack-laden process of cleaning a dataset.
    \protect\circled{B} shows the full interface for a sample of the StackOverflow dataset.  
    Each error type has a distinct color (e.g., red groups correspond to average anomalies), Upon selecting the group, \sys shows a list of wrangling/repair actions on the right and shows a visual preview of the chart after the repair.}
    \label{fig:snapshot}

\end{figure*}

\parahead{Transforming visualizations into repair interfaces. } A key technical insight of \sys is to treat visualizations as active substrates for user-driven group data transformation. By constructing index structures that link anomalies to data groups and anomaly types, \sys enables responsive, bidirectional interactions: users can trigger repairs through visual selections and observe the systemic consequences instantly. Moreover, through user-defined detector and repair functions, the system accommodates domain-specific anomalies while preserving flexibility and reproducibility through automatic code generation.

\parahead{Motivating Example.}
Consider the table in Figure~\ref{fig:snapshot}, which shows two groups from a larger dataset: \( G1 \) = \{\( \texttt{Income} \mid \texttt{Country} = \text{"Bhutan"} \)\} and \( G2 \) = \{\( \texttt{Income} \mid \texttt{Degree} = \text{"BS"} \)\}. Both groups have \texttt{Income}  anomalies, including outliers, missing values, and inconsistent data types.

Lou, a data scientist, is tasked with preparing this dataset for downstream analysis. Using a traditional workflow---such as importing the data into Python---Lou encounters several challenges: \textbf{Sparse anomalies:} Errors are scattered and infrequent, making them hard to detect. \textbf{Interdependent groups:} Fixing an anomaly in one group may unintentionally distort others. For instance, removing all zero-income rows from \( G1 \) could leave \( G2 \) with insufficient data. \textbf{Iterative debugging:} Writing and refining data cleaning scripts requires multiple trial-and-error cycles to validate correctness and completeness~\cite{schelter_tutorial}.

By contrast, if Lou uses \sys (Figure~\ref{fig:snapshot}~\circled{A}), the system automatically highlights anomalous groups for inspection. Lou is presented with wrangling suggestions specific to each error type---such as imputation, deletion, or conversion---and can apply these repairs interactively. Crucially, the visual interface reveals how each action affects related groups in real time, allowing Lou to iteratively explore, evaluate, and undo changes as needed. Once the data is in a satisfactory state, \sys can export a Python script encoding all the wrangling steps for future reuse or automation.

A prototype of \sys was demonstrated at VLDB 2025~\cite{buckaroo_demo}. The demo version operates entirely in client-side memory and is intended for smaller datasets. While suitable for showcasing interaction design, it lacks the scalability features required for real-world deployment. In this paper, we present our ongoing efforts to scale \sys by introducing server-side storage, differential snapshot management, and efficient update propagation---making the system scalable and practical for large, real-world datasets. Throughout the paper, we use the terms "anomaly" and "error" interchangeably, as well as "wrangling" and "repair."

\begin{figure*}[t]
  \centering
  \includegraphics[width=0.7\linewidth]{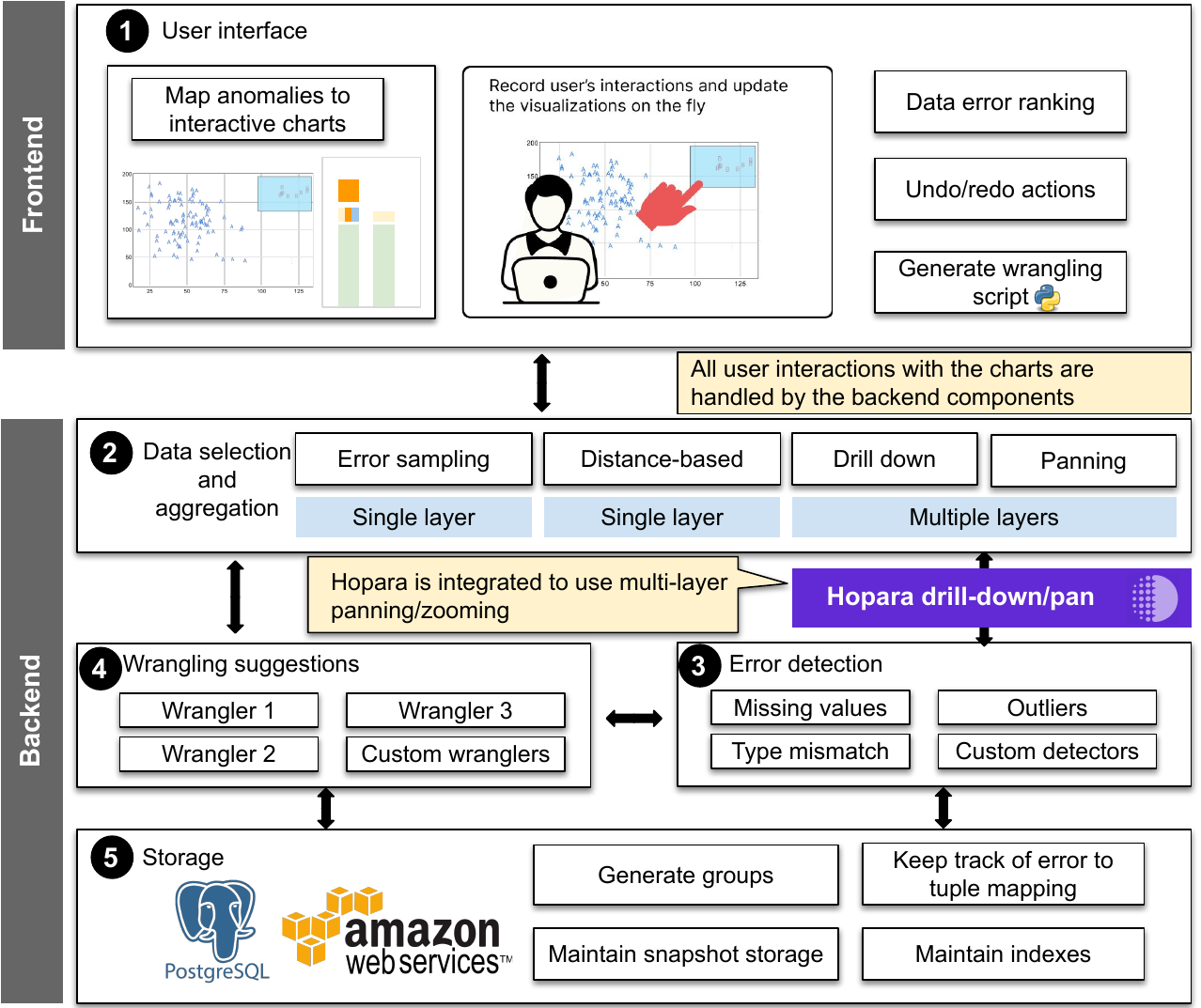}

  \caption{\sys system architecture. \protect\circled{1} The user interface visualizes anomalies and supports interactive exploration, ranking errors, undo/redo, and script generation. \protect\circled{2} The data selection and aggregation layer provides single-layer and multi-layer navigation through error sampling, distance-based sampling, drill-down, and panning. \protect\circled{3} The backend implements built-in and custom error detectors. \protect\circled{4} \sys offers built-in and user-defined wrangling functions. \protect\circled{5} The storage layer manages group generation, snapshot storage for undo/redo, error–tuple mappings, and indexes using PostgreSQL and AWS infrastructure.}
  \label{fig:architecture}

\end{figure*}

\parahead{Contributions.}
\sys introduces a new paradigm for data wrangling by tightly integrating anomaly detection, visual exploration, and repair within a single interactive interface. This paper makes the following contributions:
\begin{itemize}[leftmargin=*]
    \item A group-based abstraction that organizes anomalies into interpretable visual summaries, enabling users to interact with data through an orchestrated set of interactive charts.
    \item An extensible framework for registering custom error detectors and repair functions, allowing domain-specific wrangling logic to be incorporated.
    \item Efficient indexing and overlap-tracking structures that support localized, low-latency anomaly detection and visualization updates across interdependent views.
    \item A snapshot storage module that enables undo/redo actions and code generation, while maintaining scalability over large datasets.
\end{itemize}

\noindent{}By enabling users to \emph{see}, \emph{understand}, and \emph{repair} subgroup anomalies through a single, unified visual interface, \sys represents a paradigm shift in how practitioners wrangle data. It transforms data wrangling from a brittle and opaque task into an intuitive, transparent, and reproducible process.

\section{System Overview}

\sys is a visual data wrangling system that couples anomaly detection, visualization, and guided repair through a direct manipulation interface. Figure~\ref{fig:architecture} illustrates the overall architecture, comprising five components that span frontend interactions and backend processing. To support large datasets, \sys manages all data storage and access through a Postgres backend.

The workflow begins when a user uploads a tabular dataset through the user interface (UI) as illustrated in Figure~\ref{fig:architecture}~\circled{1}. \sys then stores the data into a Posgres database and generates groups by projecting numerical attributes onto categorical attributes (Figure~\ref{fig:architecture}~\circled{5}). The database also stores metadata linking each tuple to its associated errors.
 For each group, built-in or user-defined detectors are used to identify anomalies such as missing values, outliers, or type mismatches (Figure~\ref{fig:architecture}~\circled{3}). These anomalies are visualized through interactive charts, where users can inspect and select problematic groups. Since plotting every data point is impractical, \sys employs data selection and aggregation strategies to determine which subset of the table to visualize (Figure~\ref{fig:architecture}~\circled{2}).
 Based on the anomaly type, \sys presents corresponding wrangling suggestions---both default and user-defined---that users can apply directly to the chart (Figure~\ref{fig:architecture}~\circled{4}).

As users manipulate the data visually, the system tracks changes, re-runs localized detection only on affected groups, and updates all impacted views efficiently. All user actions are logged, and a differential snapshot mechanism ensures storage efficiency and supports undo/redo functionality (Figure~\ref{fig:architecture}~\circled{5}). \sys also creates Postgres indexes for all the attribute combinations in the charts for efficient data lookups. Finally, once the user is satisfied with the cleaned data, \sys generates an executable Python script that captures the full sequence of wrangling operations for reuse or automation. Currently, \sys only generates Python scripts, but we intend to support other target languages such as R.
We now describe the major components of \sys.

\subsection{Group Generation}

In \sys, groups serve as the fundamental abstraction for detecting and visualizing anomalies. A group is defined as a subset of the dataset obtained by projecting a numerical attribute (e.g., \texttt{Income}) onto a categorical attribute (e.g., \texttt{Country}). For example, the group \{\( \texttt{Income} \mid \texttt{Country} = \text{"Bhutan"} \)\} corresponds to the set of \texttt{Income} values for all records where the country is ``Bhutan''. This group, along with others defined by different country values, can be visualized in a chart such as a heatmap with \texttt{Country} on the X-axis and \texttt{Income} on the Y-axis. Users can control this process by selecting the projection columns and adjusting granularity (e.g., setting a minimum group size). Using group-based analysis, rather than inspecting individual rows, offers several benefits:
\begin{itemize}[leftmargin=*]
    \item \textbf{Summarization:} Groups compress many data points into coherent aggregates, making it easier to detect outliers, missing values, or irregular patterns at a glance.
    \item \textbf{Isolation within attributes:} Grouping has long been used to isolate error detection and repair (e.g., blocking~\cite{xubook} and subgroup discovery~\cite{subgroup1}). Groups defined over a single categorical attribute are disjoint---each row belongs to exactly one group per attribute. This means that repairing an anomaly in one group (e.g., \texttt{Country = Bhutan}) does not require updates to other groups using the same attribute with a different value.
\end{itemize}

\noindent{}However, as in the motivating example, groups defined over different attributes can overlap, since a single row may belong to multiple groups across multiple charts. \sys tracks these dependencies and selectively updates only affected groups when a repair is made (more details in Section~\ref{sec:localized}).

\subsection{Interactive User Interface}
\sys generates a chart matrix (see cropped view in \linebreak Figure~\ref{fig:snapshot}~\circled{B}) where data groups are represented in a heat map. Detected anomalies are visually overlaid across chart types---scatterplots, histograms, heatmaps---with groups color-coded by their dominant anomaly type. Users can interactively explore, filter, and manipulate these groups directly through the visual interface.

\sys records all user actions---such as applying a repair, exploring a group, or undoing a prior fix---and communicates them to the backend to maintain a synchronized and consistent snapshot. The UI also displays ranked anomaly (based on their frequency in the data) summaries and offers a repair kit sidebar to surface appropriate wrangling options for selected groups. The main features of the UI  are as follows:

\begin{itemize}[leftmargin=*]
    \item \textbf{Dynamic anomaly mapping:} \sys continuously overlays detected errors onto the corresponding chart elements, visually encoding their severity and type. This visual contextualization allows users to spot data anomalies  across groups at a glance and understand how errors are distributed throughout the dataset before initiating repairs.
    
    \item \textbf{Immediate feedback:} When a user applies a wrangling operation, all affected charts and summaries update instantly to reflect the modified data. This tight feedback loop helps users reason about both the local and global consequences of their actions, preventing unintended distortions and supporting rapid trial-and-error exploration.
    
    \item \textbf{Iterative editing:} Every transformation—whether a value imputation, deletion, or type correction—is logged and reversible. Users can freely undo or redo prior steps, enabling a flexible and exploratory workflow in which they can test alternative repair strategies without committing prematurely or losing previous progress.
    
    \item \textbf{Script generation:} After users reach a satisfactory data state, \sys compiles the full sequence of wrangling actions into a Python script. This script preserves provenance, supports reproducibility, and allows users to integrate their visually authored cleaning pipeline into downstream analytical workflows.
\end{itemize}

\section{Error Detection and wrangling}

\sys supports both generic and domain-specific error detection to surface data issues during interactive wrangling. For each detected error type, \sys provides corresponding wranglers that can be applied directly through the interactive charts.

\subsection{Error Detection}
Built-in detectors identify common anomalies such as missing values, outliers, type mismatches, and small groups (groups containing few points). However, data quality is often domain-dependent~\cite{xubook}, requiring customized logic. To address this, \sys offers an extensible API through which users can define their own detectors that operate at the group level, enabling flexible and reusable domain-specific validation.

\parahead{Built-in Error Types. }
\sys supports the following built-in error types:
\begin{itemize}[leftmargin=*]
    \item \textbf{Missing Values:} Identifies null or undefined cells within groups.
    \item \textbf{Outliers:} Flags values outside a configurable threshold (e.g., 2 standard deviations from the global mean).
    \item \textbf{Type Mismatches:} Detects non-numeric entries in numeric columns (e.g., “12k” in a salary field).
    \item \textbf{Group Incompleteness:} Marks groups with cardinality below a minimum threshold.
\end{itemize}

In the current \sys prototype, built-in error detectors are implemented as SQL queries, allowing them to run directly on the underlying database.

\parahead{Custom Error Detectors. } Users can register domain-specific detectors via a simple API. A detector is a function that receives a group and target attribute and returns a list of anomalous tuples. Each custom detector is mapped to a unique error code. This extensibility allows domain experts to define tailored quality checks (e.g., clinical code mismatches or sensor dropouts). For instance, the following custom detector detects if an income is less than 0:

\begin{lstlisting}[style=mypython]
def custom_detector(df: pd.DataFrame = None, 
                    target_column: str = "", 
                    error_type_code: str = "") -> list:
    if error_type_code == "negative_income":
        if df is None:
            # Write SQL query to detect the error
            query = f"SELECT id FROM salary_table WHERE {target_column} < 0"
            
            return sys.get_row_ids(query)
\end{lstlisting}

The detector returns a list of row indices corresponding to tuples exhibiting a specified error type. As shown in the listing above, certain errors—such as negative values—can be efficiently detected using a SQL query. However, not all error types are expressible in SQL~\cite{schelter_tutorial}. To accommodate such cases, \sys supports an optional Pandas DataFrame input, allowing detectors to operate directly on in-memory data when SQL alone is insufficient.

\subsection{Data Wrangling}
Once errors are detected, \sys enables users to explore and resolve them through interactive, direct manipulation of visualizations. Users can select anomalous groups or individual data points in the chart and invoke repair operations from a contextual repair suggestion toolkit. These actions are reversible and can be iteratively applied to explore their effect on the overall dataset.

\sys provides built-in wranglers for common error types and allows users to define custom wranglers via the \sys API, by mapping a user-defined function to a specific error code.

\parahead{Repair Suggestions. }
Upon selecting a group or data point with an anomaly, \sys presents a menu of repair options tailored to the error type. For instance, a missing value might prompt the user to choose between imputation (e.g., using the group mean) or row deletion, while a type mismatch might offer a conversion routine. For example, in Figure~\ref{fig:snapshot}~\circled{A}  a user selects a data anomaly, which prompts \sys to suggest appropriate wrangling actions. For each suggestion, a preview (Figure~\ref{fig:snapshot}~\circled{B}) of the intended repair is generated. Since datasets may contain a large number of errors, \sys prioritizes user attention by ranking data groups based on the number of anomalies they contain, surfacing the most erroneous groups first. Similarly, wrangling suggestions are ranked by their effectiveness—favoring repairs that resolve the anomaly with minimal side effects on other groups, i.e., minimal errors are caused for the other data groups.

\begin{figure}
    \centering
    \includegraphics[scale=0.45]{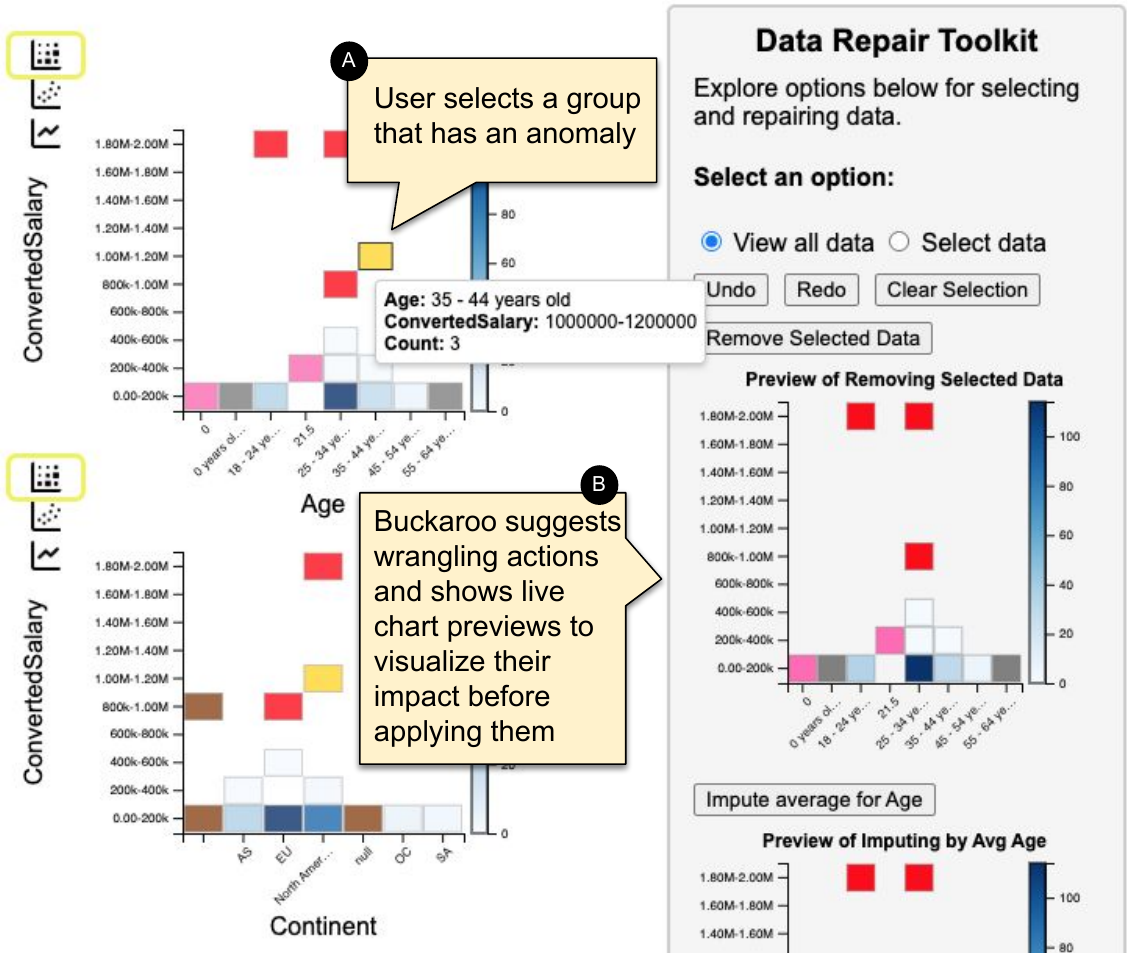}
    \caption{An overview of repairing an error through \sys's user interface. Each error type has a distinct color (e.g., red groups correspond to average anomalies), Upon selecting the group \protect\circled{A}, \sys shows a list of wrangling/repair actions on the right~\protect\circled{B} and shows a visual preview of the chart after the repair.}
    \label{fig:snapshot2}
\end{figure}

Figure~\ref{fig:snapshot2} shows an enlarged cropped view of the wrangling UI of \sys. When a user selects a problematic group (Figure~\ref{fig:snapshot2}~\circled{A}), \sys surfaces targeted wrangling options—such as imputation, deletion, or type conversion. Each candidate repair is accompanied by a live chart preview (Figure~\ref{fig:snapshot2}~\circled{B}), allowing users to assess the expected impact on the dataset before applying a change. This preview-driven interaction provides transparent, real-time feedback that makes the consequences of each action visible across related groups, helping users reason about complex group interdependencies without writing any code. By integrating visual anomaly highlighting, repair suggestion, and visual repair preview into a single loop, \sys enables data cleaning through direct manipulation of visualizations, where users can iteratively explore, refine, and undo operations with clarity and control.

\parahead{Interactive feedback.}
After a repair is applied, \sys immediately updates the visualization to reflect the modified data. Following previous architectures~\cite{tao2019kyrix}, \sys maintains a backend cache. When a data group is modified, only the affected rows in the backend cache are updated. To balance performance and persistence, \sys periodically flushes these changes to the Postgres database---by default, after every three updates, which can be configured by the user. This feedback loop allows users to observe the downstream consequences of a change, including the emergence or resolution of other errors across related groups. By visualizing repair effects, users can make informed decisions without needing to script or rerun batch detection or repair jobs.

\subsection{Localized Error Detection and Cross-Chart Dependencies}
\label{sec:localized}

A key technical challenge in \sys is ensuring that error detection remains efficient during interactive wrangling. Running anomaly detectors across the entire dataset after every repair would be prohibitively expensive and break the real-time user experience~\cite{xubook}. To address this, \sys adopts a localized and incremental error detection strategy grounded in group-based computation.

Groups defined by categorical attributes are the atomic units of visualization and error tracking. Each group is associated with a set of row identifiers (IDs). Anomaly detection is scoped to those IDs. When a repair is applied, we re-run detection only for the affected groups, avoiding unnecessary recomputation.

However, a single row can belong to multiple groups in different charts depending on the grouping attributes. A row with a missing \texttt{Income} might appear in a group under \texttt{Country=Bhutan} in one chart and under \texttt{Degree=BS} in another. A wrangling action on that row could therefore impact multiple visualizations.

 To efficiently handle such cross-chart dependencies, \sys maintains a \textit{group overlap graph}, where each node corresponds to a group and an undirected edge connects any two groups that share one or more rows. When a group is updated, \sys consults this graph to determine which groups are affected and selectively re-runs error detection only on those connected components. 

This strategy strikes a balance between responsiveness and correctness. It avoids full dataset reprocessing while preserving the accuracy of visual feedback. In practice, most wrangling actions affect only a small number of rows, making this approach highly scalable for interactive use.

\section{Navigating Data Errors Through Interactive Charts}

A central design goal of \sys is to enable users to identify and fix data errors entirely through interactions with visualizations. However, this raises an important challenge: how can we efficiently visualize large datasets---especially when most of the data is clean---and still surface rare but critical errors? To address this, \sys supports two navigation strategies: single-layer navigation, which presents an aggregated or sampled view without panning or drill-down, which is ideal for smaller datasets; and multi-layer drill-down, which enables scalable, details-on-demand exploration through interactive panning and zooming for larger datasets.

\subsection{Single-Layer Navigation}

In single-layer navigation, the goal is to expose errors within a global view of the dataset, without overwhelming the chart with excessive data points. This is nontrivial because most datasets contain a small fraction of anomalous entries, making it difficult to visually distinguish them from the bulk of the data.

\sys implements two sampling-based strategies to make errors salient in single-layer views:

\begin{itemize}
    \item \textbf{Error-First Sampling:} For each group, \sys includes all anomalous records in the chart, ensuring no error is left unvisualized. To provide context, it randomly samples a small number of non-anomalous records from the same group or surrounding groups. This preserves visual contrast while maintaining a manageable rendering cost.

    \item \textbf{Distance-Based Sampling:} In cases where context is important (e.g., for identifying borderline outliers), \sys also supports sampling based on similarity to error points. For instance, it may select points close to the error cluster in feature space to help users understand how the anomaly deviates from the norm.
\end{itemize}

These strategies allow users to rapidly scan the dataset for anomalies, explore diverse error types, and compare them against typical values—all within a single visual layer.

\subsection{Multi-Layer Navigation}

While single-layer navigation offers broad coverage, it is insufficient for large-scale or high-dimensional datasets. To support scalable exploration, \sys integrates a multi-layer navigation engine that enables users pan, and drill down into data regions of interest—loading only the relevant subsets into view.

This functionality is implemented through a close collaboration with Hopara\footnote{\url{https://hopara.io}}, whose high-performance pan-and-zoom engine has been embedded into \sys. The result is a an interaction model where users can:

\begin{itemize}
    \item \textbf{Drill Down:} Click on a region or cluster in the chart to reveal a more detailed view of the underlying data, including subgroup breakdowns and localized anomaly summaries.
    \item \textbf{Pan and Zoom:} Move across the chart space without reloading the entire dataset. This ensures that only the visible portion of the data is loaded and rendered at any given time.
\end{itemize}

The Hopara engine automatically runs SQL queries to fetch each region. Multi-layer navigation achieves two key goals: (1) it ensures that only a manageable volume of data is loaded into memory and visualized at once, improving scalability; and (2) it allows users to focus their attention on data regions of interest.

Together, these navigation strategies make \sys capable of handling large, messy datasets in a responsive manner while ensuring that anomalies—no matter how rare—remain visually accessible and actionable.

\section{Related work}

Efforts to improve data quality and make data preparation more accessible have evolved along two trajectories: (1)~Data cleaning research~\cite{xubook} has produced numerous techniques for detecting and repairing errors; however, little attention was given to user-friendly interfaces for data cleaning. (2)~Visual data wrangling techniques provide rich, interactive substrates for examining data and wrangling data visually. \sys operates at the intersection of these two lines of work, integrating the algorithmic foundations of data cleaning with direct visual manipulation to enable users to visually and scalably inspect, adjust, and steer repairs for \textit{subgroup analysis} as part of an exploratory workflow.

\subsection{Error detection and correction}  
Subgroup discovery is a well-established task in data engineering~\cite{subgroup1, bach2025usingconstraintsdiscoversparse}, focused on identifying statistically distinct and interpretable subsets within a dataset. 

A large body of work in the data management community addresses error detection~\cite{Zia16} and correction~\cite{readytodeploy, xubook}. Error detection methods fall broadly into two categories. Error-type-agnostic systems, such as HoloDetect~\cite{holodetect}, treat detection as a few-shot learning problem, leveraging rich representations and data augmentation to classify cells as clean or erroneous. In contrast, specialized detectors target specific classes of errors, including anomaly detection~\cite{lei_ad}, functional-dependency violations~\cite{rezig2021horizon}, outlier detection and summarization~\cite{lei_outlier}, and duplicate detection~\cite{dup_detection}.

Data repair techniques aim to correct erroneous or inconsistent data. Classical approaches rely on declarative rules—most notably functional dependencies (FDs) and denial constraints—to identify and repair inconsistencies~\cite{HoloClean2017, kolahi2009optrepair, rezig2021horizon}. These rule-driven systems typically formulate repairing as an optimization problem, searching for minimal data updates that satisfy a set of constraints. For example, \cite{kolahi2009optrepair} proposes an approximation to repair FD violations, while Horizon~\cite{rezig2021horizon} scales FD-based repairing by leveraging the interactions between FDs.

Beyond rule-based techniques, several systems adopt statistical or probabilistic strategies that do not aim to enforce hard integrity constraints. HoloClean~\cite{HoloClean2017}, for instance, frames repairing as probabilistic inference over a factor graph that integrates signals from constraints, co-occurrence statistics, and external data. 
Baran~\cite{baran} learns to repair data errors by unifying multiple correction models and combining their predictions through a binary ensemble classifier. RetClean and Lakefill~\cite{retclean2024, lakefill} leverage Large Language Models and data lakes to perform missing data imputation. OTClean~\cite{otclean} is a data cleaning framework that uses optimal transport to repair datasets violating conditional independence constraints while minimally altering their underlying distributions. These systems focus on automated repair generation engines but do not address user interface usability.

\sys complements this body of work by introducing a visual, interactive substrate for exploring and applying repairs. Rather than replacing existing detection or correction algorithms, \sys remains agnostic to how errors are identified or candidate fixes are produced. It instead focuses on enabling direct manipulation of data repairs through visualizations that expose anomalies, and preview the effects of alternative fixes. \sys provides a unifying interactive layer that bridges the gap between back-end repair logic and the user-facing component of data preparation.

\subsection{Visualization}  

We build on prior work in interactive data wrangling~\cite{potterwheel} and self-service data preparation~\cite{Hellerstein2018SelfServiceDP}. Wrangler~\cite{kandel2011wrangler, wrangler_cidr}—later commercialized as Trifacta—introduced a direct-manipulation paradigm in which transformation scripts are synthesized from user interactions with graphical interfaces, including visualizations. \sys follows this paradigm but shifts the focus to subgroup anomaly detection and correction. In addition, \sys introduces error-centric sampling and aggregation strategies to support scalable analysis. While Trifacta also provides sampling capabilities~\cite{alteryx_sampling_overview_2025}, available documentation suggests that these strategies are not explicitly designed around error detection. An important follow-up work is Profiler~\cite{kandel2012profiler}, which subsequently developed statistical displays of collections of errors across a dataset, a design which we draw on in \sys. \sys is complementary to Profiler in that it provides support for subgroup anomaly analysis.

Another recent work is Dango~\cite{chen2025dango} which extends visual data wrangling with natural language prompting. \citet{xiong2022revealing, xiong2022visualizing} explore techniques for visualizing wrangling scripts. \sys takes the inverse approach by embedding wrangling actions directly within visualizations. 
\citet{kasica2020table} developed a framework describing the wrangling operations available to multi-table data, specifically in data journalism contexts, which was subsequently used by \citet{xiong2022revealing}.
\citet{shrestha2021unravel} develop a method for visualizing and manipulating data frame wrangling operations governed by fluent interfaces like Pandas. 
\citet{zhu2025visegpt} explore how visualization can be used to support understanding automatically generated data wrangling scripts through a bespoke Gantt chart. 
Scorpion~\cite{wu13scorpian} draws on user-identified outliers to synthesize predicates that explain outliers. 
\sys falls into the lineage of these systems, drawing on many of the usage patterns demonstrated in those systems. Our work is differentiated from this previous ones through our focus on subgroup analysis. 

Many visual data wrangling systems can be seen as a specialized form of visualization recommendation system (for which \citet{zeng2024systematic} provide an overview), in which the gestural input to that recommendation system that is used to drive the recommendation is the wrangling process itself. Our work draws on this tradition with a more specific focus. 
For instance, Data formulator~\cite{wang2023data, wang2025data} combines AI-based natural language guidance with a closely related visualization-by-demonstration~\cite{wang2021falx} technique, which observes that preparation and mapping are interwoven components of the presentation process. 
Lux~\cite{lee2021lux} weaves visualization recommendation into notebooks, in a manner which centers a predefined collection of analysis tasks. We invert this design, by placing wrangling inside of a visualization environment, rather than placing visualizations in a data programming context. 

Our work also draws on those that seek to typify errors visually. 
For instance, \citet{mcnutt2020surfacing} enumerates a wide range of errors that occur throughout such pipelines, typifying them into various error categories.
\citet{ruddle2023tasks} focus on a subset of this process through a taxonomy of data profiling tasks and a mapping to charts that support those tasks. 
We draw on these mappings in our visualizations that surface varied errors to the end user. 

Closely related to our own work is Arachnid~\cite{shou2019arachnid} which  explores using visualizations as medium for modifying data, particularly through modifying visual representations of data. In contrast, we use visualizations as a platform for already identified errors. 
More distantly \citet{saket2019investigating} explores the use of direct manipulation as a means to specify graphical encodings. In contrast, we draw on this interaction channel as a means to provide guidance to the data cleaning system. 

Lastly, our work draws on the long history of visual analytics systems connected with data management systems. These include a wide array of different tools and systems~\cite{battle2020structured, wu2014case}. Of particular relevance are GUI-based visual analytics systems, such as Polaris~\cite{stolte2002polaris} (subsequently commercialized as Tableau). Our work draws on the patterns and traditions of these systems, but focuses on the subgroup analysis within visual data wrangling.

\section{Preliminary results \& next steps}
\sys is a work in progress and is actively being developed. The current prototype is available at \textcolor{blue}{\url{https://github.com/shape-vis/BuckarooVisualWrangler}}. Below, we present a preliminary evaluation of the system.

\subsection{Expert review}

As a basic exploration of the applicability of our design, we consulted two CTOs---one at a data integration company and the other at a data visualization company---in an expert review~\cite{harley2018ux}, in which experts were shown a prototype version of \sys.
Both agreed that the system would likely be useful for data wrangling, particularly highlighting  that many users are impeded by high-barriers on wrangling for large datasets, and that \sys had the potential to significantly lower those barriers.

However, a common concern was related to the usability of the system in the presence of large-scale datasets. They also stressed that the usability of the system will depend on how well we can summarize erroneous data on the charts as having charts with too many errors can be overwhelming. 

Additional validation of the usability  of this design, particularly for the novice user we target, is necessary future work. However, this initial review is heartening to the validity of this design: centering wrangling in a visualization-based medium seems promising.

\subsection{Runtime results}
To explore the worries expressed by our experts, we ran a set of preliminary experiments on the \sys runtime. Each experiment simulates a workload of 50 front-end wrangling operations, measuring backend processing time and frontend re-plotting latency. These experiments were run on a MacBook Pro with an Apple M4 CPU and 16 GB of RAM.

We use three datasets: StackOverFlow~\footnote{https://survey.stackoverflow.co} which has 38,091 rows and 21 columns, The Chicago Crime dataset~\footnote{https://data.cityofchicago.org/Public-Safety/Crimes-2001-to-Present/ijzp-q8t2/about_data} containing 249,542 rows and 17 columns, and the Adult Income dataset~\footnote{https://www.kaggle.com/datasets/wenruliu/adult-income-dataset} which has 48,843 rows and 15 columns. We compare \sys's performance using direct SQL queries over PostgreSQL versus relying on Pandas DataFrames for backend computation. We can clearly see from Table~\ref{tab:wrangling-runtime} that the average response time is much lower when using Postgres, and that \sys achieves a response time of less than a second for the data removal (remove a data point) and data imputation (replace value by average of column) wrangling operations.

\begin{table}[t]
\centering
\caption{Runtime comparison of wrangling operations in Postgres vs. Pandas. Across all wrangling operations, Postgres significantly outperforms Pandas.}
\resizebox{1.0\linewidth}{!}{
\begin{tabular}{|l|c|c|c|c|}
\hline
\textbf{Dataset} & \makecell{\textbf{Postgres} \\ \textbf{(removal)}} & \makecell{\textbf{Postgres} \\ \textbf{(impute)}} & \makecell{\textbf{Pandas} \\ \textbf{(removal)}} & \makecell{\textbf{Pandas} \\ \textbf{(impute)}} \\
\hline
StackOverflow & 0.18 sec & 0.16 sec & 1.69 sec & 1.27 sec \\
Adult Income & 0.15 sec & 0.13 sec & 1.40 sec & 1.17 sec \\
Chicago Crime & 0.71 sec & 0.68 sec & 5.87 sec & 5.29 sec \\
\hline
\end{tabular}
}
\vspace{5mm}
\label{tab:wrangling-runtime}
\end{table}

\parahead{Hopara evaluation.}
While full integration of \sys with Hopara is still in progress, we successfully implemented wrangling actions within a Hopara drill-down application backed by a Postgres instance on Amazon Web Services. In particular, we measured the latency of row removal triggered from an interactive Hopara bar chart. Across 20 interactions, the average response time was \textbf{173\,ms} and \textbf{201\,ms} for the Adult Income dataset, and the StackOverFlow dataset, respectively.
 
\subsection{Next Steps and Concluding Remarks }
We are finalizing the implementation of \sys and its integration with Hopara. As part of this effort, we are developing an efficient storage layer based on differential snapshots, avoiding the overhead of storing full copies after each repair. 

In conclusion, this work develops the idea of a direct manipulation-based data wrangling tool that is mediated through the graphical medium of visualization. 
We demonstrate, through our prototype \sys,  how a number of key usability features in such a system, such as undo-redo, can be integrated into such a design in an inherently scalable manner. 
A key facet of this scalability is our notions of extensibility, which allow for the construction of domain-specific error detectors and wranglers. 
Through this work, we seek to make data wrangling more approachable by shifting to a straightforward-to-understand graphical medium.

\section*{Acknowledgments}
We sincerely thank the anonymous reviewers, Michael Stonebraker, and Joseph Hellerstein for their insightful and constructive comments, which greatly improved both the project and the paper.

\bibliographystyle{ACM-Reference-Format}
\bibliography{sample-base}

\end{document}